\documentclass[11pt]{JHEP3}
\usepackage{latexsym}
\usepackage{mathrsfs}
\usepackage{amssymb}
\usepackage{amsmath}
\usepackage{graphicx}
\usepackage{cleveref}

\newcommand{\half}{\frac{\scriptstyle 1}{\scriptstyle 2}}

\newcommand{\C}{\mathbb{C}}

\newcommand{\CP}{\mathbb{CP}}

\newcommand{\R}{\mathbb{R}}

\newcommand{\p}{\partial}

\newcommand{\e}{\mathrm{e}}

\newcommand{\CI}{\mathcal{I}}

\newcommand{\cO}{\mathcal{O}}

\newcommand{\rd}{\, \mathrm{d}}

\newcommand{\1}{{\rm 1\hskip-0.25em I}}
\newcommand{\al}{\mathfrak{a}}
\newcommand{\bet}{\mathfrak{b}}

\newcommand{\Al}{{\scalebox{0.6}{$\mathfrak{A}$}}}
\newcommand{\Asmall}{{\scalebox{0.6}{$A$}}}
\newcommand{\dAsmall}{{\scalebox{0.6}{$\dot A$}}}
\newcommand{\Bet}{{\scalebox{0.6}{$\mathfrak{B}$}}}
\newcommand{\sL}{{\scalebox{0.6}{$L$}}}
\newcommand{\sR}{{\scalebox{0.6}{$R$}}}
\newcommand{\sA}{{\scalebox{0.6}{$A$}}}
\newcommand{\sB}{{\scalebox{0.6}{$B$}}}
\newcommand{\ssA}{{\scalebox{0.5}{$A$}}}
\newcommand{\ssB}{{\scalebox{0.5}{$B$}}}

\newcommand{\ssL}{{\scalebox{0.5}{$L$}}}
\newcommand{\ssR}{{\scalebox{0.5}{$R$}}}
\newcommand{\sN}{{\scalebox{0.6}{$N$}}}

\newcommand{\pf}{\mathrm{Pf}\,}

\newcommand{\be}{\begin{equation}\label}
\newcommand{\ee}{\end{equation}}
\newcommand{\bea}{\begin{eqnarray}\label}
\newcommand{\eea}{\end{eqnarray}}

\newtheorem{propn}{Proposition}

\title{Supersymmetric S-matrices  from the worldsheet in 10 \& 11d}
\author{Yvonne Geyer,\\
Department of Physics, Faculty of Science, Chulalongkorn University,
Thanon Phayathai, Pathumwan, Bangkok 10330, Thailand,\\
\& Institute for Advanced Study, 1 Einstein Drive, 08540 Princeton NJ, USA}
\author{Lionel Mason,\\
Mathematical Institute, University of Oxford, Woodstock Road, Oxford OX2 6GG, UK}

\abstract{
We obtain compact formulae for tree super-amplitudes for  10 and 11-dimensional supergravity and 10-dimensional supersymmetric Yang-Mills and Born-Infeld.   These are based on the  \emph{polarised scattering equations}.  These incorporate  polarization data into a spinor field on the Riemann sphere and arise from a twistorial representation of ambitwistor strings in 10 and 11 dimensions.  They naturally extend amplitude formulae to manifest maximal supersymmetry.  The framework is the natural generalization of twistorial ambitwistor string formulae found previously in four and six dimensions and is informally motivated from a vertex operator prescription for a family of supersymmetric  worldsheet ambitwistor string models.}
\begin{document}


\section{Introduction}
M-theory is approximated by 11d supergravity and is often characterised as the theory that provides the natural geometric backgrounds for  supersymmetric membranes.   One might therefore expect that supermembranes  should be needed to construct amplitudes for 11d supergravity \cite{Berkovits:2002uc} rather than superstrings, whose backgrounds are naturally described by supergravity theories in 10d.  However, in this paper we propose  formulae for the massless  tree-level S-matrix of 11d supergravity based on string theories in ambitwistor space, the space of complex null geodesics.   We also explain the analogous framework for 10d superamplitudes.

Ambitwistor strings \cite{Mason:2013sva,Adamo:2013tsa, 
Casali:2015vta} provide novel  formulations of massless quantum field theories that naturally generalize the 4d twistor-strings \cite{Witten:2003nn, Berkovits:2004hg,Roiban:2004yf,Cachazo:2012pz,Skinner:2013xp}.
They directly yield the remarkable formulae of Cachazo, He and Yuan (CHY), that express tree-level amplitudes as integrals over the moduli space of marked Riemann spheres, that  localize on solutions to the \emph{scattering equations} \cite{Cachazo:2013hca,Cachazo:2014xea}.  However, the CHY formulae do not naturally manifest  supersymmetry. Fermionic amplitudes are accessible from the Ramond sector of the ambitwistor string \cite{Adamo:2013tsa} and the pure spinor ambitwistor string  \cite{Berkovits:2013xba,Gomez:2013wza}  manifests supersymmetry, but it remains difficult to generate explicit closed-form formulae beyond four points.

In 4d \cite{Geyer:2014fka}\footnote{For the 4d case, see also  \cite{Witten:2004cp} for analagous formulae arising from the original twistor-string;  they have twice as many delta functions and more moduli, but  are shown to be equivalent to the 4d ambitwistor formulation in \S5.2.2. of \cite{Geyer:2016nsh}.} 
 and 6d \cite{Geyer:2018xgb}, this was remedied by working in a twistorial representation of the model. This naturally manifests supersymmetry giving rise to compact formulae for superamplitudes, manifesting supersymmetry,   
now localizing on the \emph{polarized scattering equations} that  extend the scattering equations to incorporate polarization data.

Here we give the natural extension of these ideas to 10 and 11  dimensions, and   present the full, manifestly supersymmetric S-matrix for 11d supergravity and a variety of theories in  10d.
Some ingredients have already been presented in the literature: the \emph{tiny group} that leads to the definition of supermomenta in \cite{Boels:2012ie} and its links to ambitwistor-strings in twistor coordinates in \cite{Bandos:2014lja}. 
 The formulae are again localized on the polarized scattering equations.
We give the basic structure of ambitwistor string vertex operators in these coordinates and show how they lead to polarized scattering equations in 10 and 11 dimensions.  Although we do not give a complete quantization of these models, the structures we obtain provide the necessary ingredients for supersymmetric amplitude formulae. We first set out the 11 dimensional framework for M-theory amplitudes, then the corresponding formulae in 10 dimensions, and explain how to reduce to four dimensions to make contact with \cite{Geyer:2014fka} providing a proof at least for the lower lying formulae.

\section{11d supergravity}
\paragraph{Little groups and tiny groups.}
In $d$-dimensions, the little group is  $\mathrm{SO}(d-2)\subset \mathrm{SO}(d)$ inside the stabilizer of a null momentum vector, $k_\mu$,  $\mu=0,\ldots,d-1$. Polarization states for massless particles are  representations of this little group.  
 Let $\Gamma_\mu$ denote the Clifford matrices, then the null condition  gives  $(k\cdot \Gamma)^2=0$.  It is a standard result  that the kernel of $k\cdot \Gamma$ is half the dimension of the spin space and can be identified with the spin space of the little group.  These little group spinors give, for example, the polarization states for the massless chiral Dirac equation of momentum $k$.

In 11d, the spin space is 32 dimensional indexed by $\al,\bet =1,\ldots 32$, and  spinor indices can be raised and lowered with  a skew form $\varepsilon_{\al\bet}$.  The kernel of $k\cdot \Gamma$, the spin space for the little group, can be indexed by $\alpha, \beta=1\ldots 16$ that can be raised and lowered with a symmetric form $\delta^{\alpha\beta}$. We introduce the basis $\kappa_{\al\alpha}$  of the kernel of $k\cdot \Gamma$ normalized by \footnote{We follow the conventions of Penrose \& Rindler \cite{Penrose:1986ca} for spinors in higher dimensions.}
\begin{equation}
 \kappa_{\al \alpha}\kappa_{\bet}^{\alpha}=\Gamma_{\al\bet}^\mu k_\mu\,, \quad \, \Gamma^{\al\bet}_\mu \kappa_{\al \alpha}\kappa_{\bet \beta}=-2 k_\mu \delta_{\alpha\beta}.\label{11d-norm}
\end{equation}
We take gluon polarization data  to be null vectors $e_\mu$ with $k\cdot e=0$.   With respect to such a choice, the \emph{tiny group} \cite{Boels:2012ie} is the (now complex)  $\mathrm{SO}(d-4)$ inside the stabilizer of both $k_\mu$ and $e_\mu$. In such a situation we will have a common kernel to $k\cdot \Gamma$ and $e\cdot \Gamma$ as
\begin{equation}
\{e\cdot \Gamma,k\cdot\Gamma\}=k\cdot e\1 =0.
\end{equation}
This joint kernel can be identified with the 8 dimensional spin space of the tiny group, indexed by $a=1,\ldots , 8$, and we represent its basis by
$\epsilon_{\al a}=\kappa_{\al \alpha}\epsilon^\alpha_a\, .
$ 
When $e$ and $k$ are linearly independent, these satisfy the important (semi-) purity  relations 
\begin{equation}
\Gamma_\mu^{\al\bet}\epsilon_{\al a}\epsilon_{\bet b}=2k_\mu \epsilon^\alpha_a\epsilon_{\alpha b}=0\, .\label{semipure}
\end{equation}
This follows from  using \eqref{11d-norm} and its analogue for $e_\mu$ to see that $\Gamma^{\al\bet}_\mu \epsilon_{\al a}\epsilon_{\bet b}$ is proportional to both $k_\mu$ and $e_\mu$, and so must vanish.  We also impose the normalizations 
\begin{equation}\label{2form-norms}
\epsilon_{\al a}\epsilon_{\bet }^a=\Gamma^{2\,\mu\nu}_{\al\bet}e_\mu k_\nu,\quad \Gamma^{2\,\al\bet}_{\mu\nu}\epsilon_{\al a}\epsilon_{\bet b}=-8\delta_{ab}e_{[\mu}k_{\nu]},  
\end{equation}
where $\Gamma^{2}_{\mu\nu}=\Gamma_{[\mu}\Gamma_{\nu]}$ etc.\ as usual.

\paragraph{The polarized scattering equations.}
We take gravity polarization data to be  metric perturbations of the form $\delta g_{\mu\nu}=e_{\mu} e_{\nu} \e^{ik\cdot x}$ where $e_{\mu}$ is null, or equivalently $\epsilon_{\al a}$ or $\epsilon^\alpha_{a}$ satisfying \eqref{semipure} and \eqref{2form-norms}.

The scattering equations associate $n$ points $\sigma_i$ on the Riemann sphere to $n$ null momenta $k_{i\mu}\in \R^d$, $i=1,\ldots n$, subject to momentum conservation $\sum_i k_i=0$.  First introduce the meromorphic, M\"obius-invariant one-form
\begin{equation}
\label{rns-mom}
P_\mu(\sigma)=\sum_i\frac{k_{i\mu}}{\sigma-\sigma_i}\rd\sigma\, .
\end{equation}
The scattering equations are  $n$ equations on the $\sigma_i$, encoding $P^2=0$ for all $\sigma$:
\begin{equation}
\label{SE}
\underset{\sigma_i}{\mathrm{Res}}\, \frac{1}{2}P^2(\sigma)=k_i\cdot P(\sigma_i)=\sum_j \frac{k_i\cdot k_j}{\sigma_{ij}}=0\, .
\end{equation}
Since $P$ is null, we can hope to find $\lambda_\alpha^\al(\sigma)$, satisfying analogues of \eqref{11d-norm},
\begin{equation}
\lambda_{\alpha\al}\lambda_\bet^\alpha= \Gamma_{\al\bet}^\mu P_\mu\, , \quad \Gamma_\mu^{\al\bet}\lambda^\alpha_\al\lambda^\beta_\bet=-2P_{\mu} \delta^{\alpha\beta}\,.
\label{normP11} 
\end{equation}
Since  $k\cdot P=0$, we can again apply the tiny group argument now to $k_i$ and $P(\sigma)$ near $\sigma_i$, leading to a joint 8-dimensional kernel of $k\cdot \Gamma$ and $P\cdot \Gamma$.  This kernel is spanned  by a pair of $8\times 16$ matrices  $(u_{a\alpha},v_{a\alpha})$ 
subject to the \emph{polarized scattering equations}
\begin{equation}
 u_{ia\alpha}\lambda^\alpha_\al(\sigma_i) =v_{ia\alpha}\kappa_{i\al}^\alpha\, .
 \label{SE11}
\end{equation}
Note that the $\alpha$-indices on the $v_{ia\alpha}$s are those for the little group for $k_{i\mu}$ whereas that on the $u_{ia\alpha}$ are global little group indices associated to $P_\mu$. 
The variables $(u_{a\alpha},v_{a\alpha})$ are defined up to a $\mathrm{GL}(8)$-transformation of the $a$-indices, and  satisfy
 \begin{equation}
u_{a\alpha}u_{b\beta}\delta^{\alpha\beta}=0\, , \quad v_{a\alpha}v_{b\beta}\delta^{\alpha\beta}=0\, ,\label{little-pure}
\end{equation}
so that these subspaces are (semi-)pure.  We have the freedom to further normalize  against $\epsilon_{\alpha a} $ by
\begin{equation}
\epsilon_{\alpha a}v^\alpha _b= \delta_{ab}\, .\label{norm-v}
\end{equation}
This reduces the GL$(8)$ freedom on the $a$-index down to SO$(8)$.\footnote{Further normalization  can be done to reduce to the  Spin$(7)$ tiny group but we wont use that in the following.}

Equation  \eqref{normP11} implies that $\lambda_{\alpha \al}$ is  a worldsheet spinor. Motivated by the ambitwistor-string model introduced later, we make the Ansatz
\begin{equation}
\lambda_{\al \alpha}(\sigma)=\sum_{i=1}^n \frac{u_{i\alpha a} \epsilon_{i\al}^a}{\sigma-\sigma_i}\sqrt{d\sigma}\, ,\label{lambda-sigma11}
\end{equation}  
where $\epsilon_{i\al}^a$ is the polarization data for the $i$th particle. 

It is a key fact that for each solution to the scattering equations $k_i\cdot P(\sigma_i)=0$, with momenta and polarization data in general position, there exists a unique $\lambda_{\al \alpha}$ satisfying \eqref{SE11}  and \eqref{lambda-sigma11}, \cite{wip}.  Briefly, this follows from a degree count of the subbundle $E\subset \mathbb{S}_\al$ where $\mathbb{S}_\al$ is the trivial bundle of spinors over $\CP^1$ and $E$ the subbundle that is annihilated by $P^m\gamma_m^{\al\bet}$.  For each index $\alpha$, $\lambda^\al_\alpha$ is a section of $E\otimes \cO(-1)$. It follows from the defining exact sequences that $E\otimes \cO(n-1)$, the bundle in which $\lambda^\al_\alpha\prod_{i=1}^n (\sigma-\sigma_i)$ takes its values, has degree $8n$. However, the ansatz \eqref{lambda-sigma11} imposes $8$ conditions per marked point thus reducing the degree to zero. Thus this 16 dimensional bundle is generically trivial with 16 sections.  These can then be normalized to satisfy \eqref{normP11}.

\paragraph{Supersymmetry and the tiny group.}
The tiny group was  introduced in  \cite{Boels:2012ie} to define  supermomenta in higher dimensions, and it was argued that there are natural choices for the ambitwistor-string  in \cite{Bandos:2014lja}. Although there the proposed reduction arises from the null $P(\sigma)$ at $\sigma_i$, but in our context this is a pole with residue $k_i$, which is not independent of $k_i$, and so does not work directly.  
We can nevertheless use a variant to  introduce supermomenta in the context of our polarized scattering equations \eqref{SE11} as follows. 
On a momentum eigenstate, the supersymmetry generators satisfy $\{Q_\al ,Q_\bet\}=\Gamma_{\al\bet}^\mu k_\mu$. This allows us to define (little-group) 
$Q_\alpha$ via 
$$
Q_\al=\kappa_\al^\alpha Q_\alpha\, ,
\quad\mbox{ satisfying  } \quad \{Q_\alpha,Q_\beta\}=\delta_{\alpha\beta}\, .
$$
The introduction of supermomenta requires the choice of an anticommuting 8 dimensional subspace of the 16  $Q_\alpha$'s.  For us, a natural choice  arises from the polarization data and solution to the polarized scattering equations   $(\epsilon_{i\alpha}^a,v_{i\alpha}^a)$ as these satisfy $ v_{i\alpha a}v^\alpha_{ib}=0=\epsilon_{i\alpha a}\epsilon^\alpha_{ib}, \epsilon_{i\alpha a} v^\alpha_{ ib}= \delta_{ab}
$. 
This however would lead to a supersymmetry representation that depends  on the solution to the polarized scattering equations via $v_i$. Instead, we choose one additional basis spinor $\xi_{i a}^\alpha$ for each particle, such that $(\epsilon_{i\alpha}^a,\xi_{i\alpha}^a)$ satisfy $ \xi_{i\alpha a}\xi^\alpha_{ib}=0$ and $ \epsilon_{i\alpha a} \xi^\alpha_{ ib}= \delta_{ab}$.
We can then define  fermionic supermomenta $q_i^a$ by the relations
\begin{equation}\label{eq:gen}
Q_{i\alpha}=\xi_{i\alpha a}q_i^a +\epsilon_{i\alpha}^a\frac{\p}{\p q_i^a}\, .
\end{equation}
The 11d supergravity massless multiplet consists of the triplet
$(h_{\mu\nu}, C_{\mu\nu\rho},\psi_\mu^\bet)$, containing a metric, 3-form potential and Rarita-Schwinger field.
The full supermultiplet is then generated from the pure graviton state $(e_\mu e_\nu,0,0)$ at $q_i^a=0$.
At $O(q^a)$ we see the 8 components of the Rarita-Schwinger field $\psi_\mu^\al=e_\mu\epsilon^\al_a q^a$ and at $O(q^aq^b)$ the 3-form $C_{\mu\nu\rho}=\Gamma_{\mu\nu\rho}^{\al\bet}\epsilon_\al^a\epsilon_\bet^bq_aq_b$ and so-on (see \S\ref{sec10d} for full details of the 10d analogues).

We define the total supersymmetry generator for $n$ particles by
\begin{equation}
Q_\al=\sum_i \kappa_{i\al}^\alpha Q_{i\alpha}=\sum_i \kappa_{i\al}^\alpha \left(\xi_{i\alpha a}q^a_i+ \epsilon^a_{i\alpha}\frac{\p}{\p q_i^a}\right).
\end{equation}

A clear consistency requirement on supergravity amplitudes is that they must be annihilated by  $Q_\al$.
We will see that  the total dependence of the supergravity superamplitude on the supermomenta in this representation should take the form of an exponential factor $\e^F$, with 
\begin{equation}
F=\sum_{i<j}^n\frac{u_{ia\alpha}u^\alpha_{jb}}{\sigma_{ij}}q_i^aq_j^b - \frac1{2}\sum_{i=1}^n \xi_{ia\alpha}v_{ib}^\alpha\;q_i^a q_i^b\, .
\end{equation}
We  discuss the origin of this factor from a worldsheet model in the next section. Supermomentum conservation is then easily verified, 
\begin{equation}
Q_\al \e^F 
=\left(\sum_i \kappa_{i\al}^\alpha  v_{i\alpha a}q_i^a- \sum_{j} \lambda_{\al\alpha}(\sigma_j)u^\alpha_{jb}q_j^b \right)\e^F =0\,,\nonumber
\end{equation}
with the second equality following from the polarized scattering equations \eqref{SE11}. This guarantees invariance under supersymmetry provided the $q$-dependence is encoded in the exponential $\e^F$.

\paragraph{11d SUGRA amplitudes.}
Our amplitude formulae take the form
\begin{equation}
\mathcal{M}_n=\int_{\mathfrak{M}_{0,n}} \rd\mu_{\scalebox{0.6}{CHY}} \,\mathcal{I}_n \label{CHY}\,,
\end{equation}
where the CHY measure on the moduli space $\mathfrak{M}_{0,n}$ of $n$ points $\sigma_i$ on $\CP^1$ is given by 
\begin{equation}
\rd\mu_{\scalebox{0.6}{CHY}}:=\frac{ \prod_{i=1}^n \bar\delta(k_i\cdot P(\sigma_i))d\sigma_i}{\mathrm{vol} (\mathrm{SL}(2)\times \C^3)}\,,
\end{equation}
with the M\"obius transformation quotient defined via the usual Faddeev-Popov methods and the $\C^3$ quotient leading to the removal of three $\bar\delta$-functions and a further Faddeev-Popov factor \cite{Mason:2013sva}.
For 11d supergravity our formula arises simply from
\begin{equation}\label{eq:M-th-ampl}
\CI_n=\det{}' M\, \e^F\,,
\end{equation}
where $M= \begin{pmatrix}
A&C\\-C^t&B
\end{pmatrix}$ is the $2n\times 2n$ CHY matrix constructed from our polarization data, 
\begin{equation}
A_{ij}=\frac{k_i\cdot k_j}{\sigma_{ij}},\; B_{ij}=\frac{e_i\cdot e_j}{\sigma_{ij}},\;\, C_{ij}=\begin{cases} \frac{e_i\cdot k_j}{\sigma_{ij}} ,& i\neq j\\
-e_i\cdot P(\sigma_i), & i=j\, ,\label{CHY-M}
\end{cases}
\end{equation}
with $\sigma_{ij}=\sigma_i-\sigma_j$. The reduced determinant is defined as $\det{}'M=\det M_{[ij]}/\sigma_{ij}^2$, where $M_{[ij]}$ is $M$ with rows and columns $i,j$ removed. 

For $q_{ia}=0$, it is clear that our formula reduces to the standard CHY formula for gravity amplitudes.  Thus  our work provides a natural supersymmetric extension to provide the full 11d supergravity multiplet.

\paragraph{Worldsheet model and vertex operators.}\label{sec:WS-model}
To motivate the polarized scattering quations and  supersymmetry factors, we introduce here a twistorial ambitwistor string model. A full quantum description of the model is beyond the scope of this letter.

Let us work in an 11d superspace with coordinates $(x^\mu,\theta^\al)$.  The  Green-Schwartz  ambitwistor-string  for supergravity \cite{Mason:2013sva} has the worldsheet action
\begin{equation}
S=\int_\Sigma P_\mu(\bar\p X^\mu -i\theta^\al \bar\p \theta^\bet \Gamma_{\al\bet}^\mu )+ \frac{e}{2}\,P^2\, , 
\end{equation}
Following \cite{Bandos:2014lja}, we solve the $P^2=0$ constraint by  $2P_\mu\delta^{\alpha\beta}=\lambda_\al^\alpha \lambda_\bet^\beta\Gamma^{\al \bet}_\mu$.  We introduce twistors $Z^\Al=(\lambda_\al,\mu^\bet,\eta)$, $\scalebox{0.8}{$\mathfrak{A}$}=1,\ldots ,64|0$ as a supersymmetric extension of spinors for the conformal group $\mathrm{SO}(13)$ \cite{Penrose:1986ca}, with a skew inner product \begin{equation}
\varepsilon_{\Al\Bet}Z_1^\Al Z_2^\Bet=\lambda_{1\al}\mu_2^\al- \mu_1^\al\lambda_{2\al} +\eta_1\eta_2\,,
\end{equation}
that can be used to raise and lower indices. The spinorial representation of the ambitwistor string is then formulated using  16  such twistors $Z^\Al_\alpha$,  related to space-time via  the incidence relations
\begin{equation}\label{11inc}
 \left(\mu_{\alpha}^\al, \eta_\alpha\right)
= \left(\frac{{\scriptstyle 1}}{{\scriptstyle 32}}\lambda_{\alpha \bet}\left(X^\mu\Gamma_\mu^{\al \bet}-16i\theta^\al \theta^\bet\right), \lambda_{\alpha \al}\theta^\al\right) . \nonumber
\end{equation}
Using $32P_\mu=\lambda_\al^\alpha\lambda_{\bet\alpha}\Gamma^{\al\bet}_\mu$, the Green Schwarz action transforms in twistor coordinates to 
\begin{equation}
S=\int_\Sigma Z^\Al_\alpha  \bar \p Z^\alpha_\Al + A^{\alpha\beta}_{M}\Gamma^{M}_{\Al\Bet}Z^\Al_\alpha Z^\Bet_\beta 
 \label{S11}
\end{equation}
Here the $Z^\Al_\alpha$  are worldsheet spinors, $A^{\alpha\beta}_{M}$ are  $(0,1)$-form gauge fields  on the worldsheet,
 and $\Gamma^{M}_{\Al\Bet}$ are the $\mathrm{SO}(13)$ gamma matrices. The $A^{\alpha\beta}_{M}$  are Lagrange multipliers for the 13d semi-purity constraints $\Gamma^{M}_{\Al\Bet}Z^\Al_\alpha Z^\Bet_\beta=0$ 
that follow from  the existence of $(x^\mu,\theta^\al,P_\mu)$ such that the incidence relations \eqref{11inc} hold \cite{wip}.

The vertex operators for this model need to reduce to $\bar\delta(k\cdot P)\, \e^{ik\cdot x}$ in the bosonic case.  Including supermomenta $q^a$, we propose\footnote{Note here that the weight of $u^\alpha_a$ as a worldsheet co-spinor cancels that of $(\mu^\al_a,\eta_\alpha)$.}
\begin{equation*}
V=\int \delta(k\cdot P)\, w \exp \left(\mu^\al_\alpha \epsilon_\al^a u_a^\alpha + \eta_\alpha u^\alpha_a q^a-\frac1{2}\xi_{a\alpha}v_{b}^\alpha\;q^a q^b\right)  \, .\label{VO11}
\end{equation*}
Here $w$ is an additional worldsheet operator depending on the polarization data whose correlators provide the determinant $\det{}'M$  as in \cite{Mason:2013sva, Casali:2015vta}. This reduces correctly to the bosonic case: using the unique solution  $(u,v)$ to the polarized scattering equations and the incidence relations together with \eqref{eq:lambda-eta-sols}, the argument of the exponential becomes  $k\cdot x+ \theta^\al\theta^\bet k_{\al\bet} +\theta^\al\kappa_\al^\alpha \xi_{\alpha a} q^a$ as appropriate for a supermomentum eigenstate.


Consider now a path-integral  with $n$ vertex operators. The exponentials in the vertex operators can  then be taken into the action,  providing sources 
$(\epsilon^a_{i\al} u_{ia}^\alpha, u^\alpha_{ia}q^a_i)$ in the equations of motion for $(\lambda_\al^\alpha,\eta^\alpha)$
\begin{equation}
\bar\p (\lambda^\alpha_\al, 2\eta^\alpha)= \sum_i (\epsilon^a_{i\al} u_{ia}^\alpha, u_{ia}^\alpha q_i^a)\bar\delta(\sigma-\sigma_i) \, , \end{equation}
The path integral then localizes onto the classical solution
\begin{equation}\label{eq:lambda-eta-sols}
\big(\lambda^\alpha_\al(\sigma),\eta^\alpha(\sigma)\big)=\sum_i \left(\frac{\epsilon^a_{i\al} u_{ia}^\alpha}{\sigma-\sigma_i} ,\frac{u^\alpha_{ia}q_i^a}{2(\sigma -\sigma_i)}\right)\, ,
\end{equation}
yielding \eqref{lambda-sigma11} as promised.  Furthermore, 
localising on these classical solution with $\mu^\al_\alpha=0$ leads to the exponential factor in the fermions
\begin{align}
\exp\left(\sum_i \eta_\alpha(\sigma_i)u^\alpha_{ia}q_i^a-\frac{1}{2}\xi_{ia\alpha}v_{ib}^\alpha\;q_i^a q_i^b\right)&=e^F\,,\nonumber 
\end{align}
giving the exponential supermomentum factor  introduced earlier.

\section{10d superamplitudes}\label{sec10d}
Much of the analysis in 11d extends straightforwardly to 10d, both by analogy and dimensional reduction.  We redefine the space-time and little-group indices to $\mu=1,\ldots 10$ and $m=1,\ldots ,8$, but maintain our spinor conventions. Note that there is no metric on the 10d spin space  $\alpha,\beta=1,\ldots ,16$. 
The little group is now $\mathrm{SO}(8)$  with two types of chiral spinor indices $a=1,\ldots ,8$ and $\dot a=\dot 1,\ldots,\dot 8$.   The Clifford matrices $\Gamma$ decompose into chiral Pauli matrices $\gamma_{\mu\alpha\beta}$, $\gamma^{\alpha\beta}_\mu$.

\paragraph{Little and tiny groups  in 10d.}
Denote the basis of the kernel of $k\cdot \gamma_{\alpha\beta}$ by $\kappa^\alpha_a$, normalized by 
\begin{equation}
\kappa_{a\alpha}\kappa_\beta^a= \gamma_{\alpha\beta}^\mu k_\mu\,,\quad \gamma_\mu^{\alpha\beta}\kappa_{\alpha}^a\kappa_{\beta}^b=-2k_{\mu} \delta^{ab}\,, 
\label{norm} 
\end{equation}
with similar dotted versions for $\kappa^\alpha_{\dot a}$.  
For null polarization vectors $e_\mu$, the joint kernel of $k\cdot \gamma$ and $e\cdot \gamma$ is now 4-dimensional in each chiral spin space, 
\begin{equation}
\epsilon _{\alpha }^A=\kappa_\alpha^a\epsilon_a^A\, , \quad \epsilon^\alpha_A=\kappa^{\dot a\alpha}\epsilon_{\dot a A}\, , \quad A=1,\ldots 4
\end{equation}
where $A$ is a spinor index for the $\mathrm{SO}(6)$ tiny group. As in 11d, we impose the normalizations
\begin{equation}
   \epsilon_{\alpha}^A\epsilon^{\beta}_A= (\gamma^{ \mu\nu})_\alpha^\beta\,e_\mu k_\nu\, ,\quad \epsilon_{\alpha}^A\epsilon^{\beta}_A\,(\gamma_{\mu\nu})^{\alpha }_{\beta}= -8\,\delta_A^B \, e_{[\mu}k_{\nu]} \, \label{pol10}.
\end{equation}
We now have full purity conditions 
\begin{equation}
\epsilon_{a}^A\epsilon^{aB}=0,  \qquad  \epsilon^{\dot a}_A\epsilon_{\dot aB}=0,\label{little-pure0}
\end{equation}
following as before because $ \gamma_\mu^{\alpha\beta}\epsilon_{\alpha}^A \epsilon_{\beta}^B$ is proportional to both $k_\mu $ and $e_\mu$ and so must vanish.

\paragraph{The polarized scattering equations.}
On the scattering equations, we decompose $P(\sigma)$ again into spinors  $\lambda_\alpha^a$ via
\begin{equation}
\lambda_{a\alpha}\lambda_\beta^a= \gamma_{\alpha\beta}^\mu P_\mu\,, \qquad \gamma_\mu^{\alpha\beta}\lambda_\alpha^a\lambda_\beta^b=-2P_{\mu} \delta^{ab}\,, 
\label{normP} 
\end{equation}
together with a similarly normalized $\lambda^\alpha_{\dot a}$. Since this is again a worlsheet spinor, we take
\begin{equation}
\lambda_{a \alpha}(\sigma)=\sum_{i=1}^n \frac{u_{iaA} \epsilon_{i\alpha}^A}{\sigma-\sigma_i}\, .\label{lambda-sigma}
\end{equation}  
where $\epsilon_{i\alpha}^A$ is the polarization data for the $i$th particle.
As before, the scattering equations $k\cdot P=0$ ensure that $k\cdot \gamma$ and $P\cdot \gamma$ share a  4-dimensional kernel,  parametrized by a pair of $4\times 8$ matrices  $(u_{aA},v_{aA})$. 
These are again subject to the \emph{polarized scattering equations},
\begin{equation}
 u_{iaA}\lambda^a_\alpha(\sigma_i) =v_{iaA}\kappa_{i\alpha}^a\, , 
 \label{SE10}
\end{equation}
and similarly $u_{i}^{\dot aA}\lambda_{\dot a}^\alpha(\sigma_i) =v_{i}^{\dot aA}\kappa_{i\dot a}^\alpha$ for the opposite chirality.
The purity conditions 
 \begin{equation}
u_{aA}u_{bB}\delta^{ab}=0\, , \qquad v_{aA}v_{bB}\delta^{ab}=0\, ,\label{little-pure10}
\end{equation}
ensure that these subspaces are totally null.  
Moreover, they are dual to the 4-space defined by the polarization data due to  the normalization
$v_{A}^a\epsilon_a^B=\delta_A^B $, giving a unique tiny group for each particle.
As in 11d, there exists a  unique solution $(u_{aA},v_{aA})$ for each solution $\{\sigma_i\}$ to the scattering equations.

\noindent

\paragraph{Supermomenta for Yang-Mills theory.}
For super Yang-Mills theory, the supersymmetry generators $Q_\alpha$ act on the supermultiplet by
\begin{equation}
 Q_\beta (e_{\mu},\zeta^\alpha)= \left(\frac{1}{2}\gamma_{\alpha\beta\mu}\zeta^\alpha, \gamma^{\mu\nu \alpha}_\beta e_{[\mu}k_{\nu]}\right).\label{10d-susy-gen}
\end{equation}
These reduce to the little group data $(e_m,\zeta^{\dot a}\!=\kappa^\alpha_{\dot a}\zeta^{\dot a})$ by $Q_\alpha\!\!=\kappa_\alpha^aQ_a$ where $\{Q_a,Q_b\}=\delta_{ab}$ acting by
\begin{equation}
Q_a(e_m,\zeta^{\dot a})=\left(-\frac{1}{2}\gamma_{m a\dot a}\zeta^{\dot a} ,\gamma^{m\dot a}_a e_m\right)\, ,\label{8d-susy-gen}
\end{equation}
where $\gamma_{ma\dot a}$ are 8d gamma matrices that relate the polarization data  $e_m$  to  $\epsilon_a^A$, $\epsilon^{\dot a}_A$ by
\begin{equation}\label{eq:e_m}
 e_m\gamma^m_{a\dot a}=\epsilon_{aA}\epsilon^A_{\dot a}\, ,\quad \gamma_{ma\dot a} \epsilon^a_A \epsilon^{\dot aB}=-2e_m\delta_A^B\, , 
\end{equation}
To construct the supersymmetry representation, we again introduce additional little group spinors $\xi_{iA}^a$ and $\xi_{i\dot a }^A$ for each particle, such that
\begin{equation}\label{eq:xi_10d}
 \xi_{aA}\xi_{bB}\delta^{ab}=0\,,\qquad \xi_A^a\epsilon_a^B=\delta^B_A\,,
\end{equation}
and similarly for $\xi_{\dot a }^A$. The 8d vector $\xi_m$ relates to $\xi^a_A$ and $\xi_{\dot a}^A$ via  the analogous relations to \eqref{eq:e_m}, and  
the $\xi_{A}^a$ and $\epsilon_A^{\dot a}$ further determine 6d
$\gamma$-matrices  by 
\begin{equation}
\gamma^{(6)}_{m AB}:=\gamma_{ma\dot a}\epsilon^{\dot a}_{[A}\xi^a_{B]}=\half \varepsilon_{ABCD} \gamma_{ma\dot a}\xi^{\dot a C}\epsilon^{a D}.
\end{equation}
We use the polarization data and the solutions to the scattering equations  to  parametrize the super Yang-Mills  multiplet,
\begin{equation}
(e_m,\xi^{\dot a})=(q^4 \xi_m+ 2\gamma^6_{mAB}q^A q^B +e_m, \epsilon^{\dot a}_Aq^A+ \xi^{\dot aA}q^3_A)\,. 
\end{equation}
Here, $q^A$ are fermionic supermomenta, with $q^4=\frac{1}{4!}\varepsilon_{ABCD}q^A \ldots q^D$ and $q^3_A=\frac{1}{3!}\varepsilon_{ABCD}q^B q^C q^D$.  On these representatives, the supersymmetry generators take the now-familiar form  
\begin{equation}
Q_{ia}=\xi_{iaA}q_i^A + \epsilon_{ia}^A\frac{\p}{\p q_i^A}\, .
\end{equation}
The full supermultiplet $(e_{[\mu}k_{\nu]},\zeta^\alpha)$  is then given by
\begin{equation}
 \left(\!\gamma_{\mu\nu \beta}^\alpha(\xi_{\alpha A}\xi^{\beta A}q^4 +2\xi_{\alpha A}\epsilon_{B}^\beta q^Aq^B + \epsilon_{\alpha }^A\epsilon^{\beta }_A ),\,\epsilon^{\alpha}_Aq^A\!+\!\xi^{\alpha A}q^3_A
\right) , 
\end{equation}
where $(\xi_{\alpha A},\xi^{\alpha A})=(\xi_A^a\kappa_{a\alpha},\xi^{\dot a A}\kappa_{\dot a }^\alpha)$, with  supersymmetry generators 
\begin{equation}
Q_{i\alpha} =\xi_{i\alpha A}q^A_i + \epsilon_{i\alpha}^A\frac{\p}{\p q_i^A}\, .
\end{equation}
For $n$ superparticles, the total supersymmetry generator is again given by $Q_\alpha=\sum_i Q_{i\alpha}$.
Motivated  by the ambitwistor string model, we define $\eta_a(\sigma)$ in analogy to \eqref{eq:lambda-eta-sols}. 
Super Yang-Mills amplitudes then only depend on the supermomenta $q^A$ via an exponential factor  $\e^{F_1}$, where 
\begin{align}
\label{susy-factor}
F_1&:=\sum_{i=1}^n \eta_a(\sigma_i) u_{iA}^aq_i^A-\frac1{2}\sum_{i=1}^n\xi_{iaA}v_{iB}^a\;q_i^Aq_j^B\nonumber\\
&=\sum_{i<j} \frac{u^a_{iA}  u_{jBa}}{\sigma_{ij}}\,q^A_iq_j^B-\frac1{2}\sum_{i=1}^n\xi_{iaA}v_{iB}^a\;q_i^Aq_j^B\,.
\end{align}
In 10 dimensions, we can extend the supersymmetry to $\mathcal{N}=2$ with an $\mathrm{SO}(2)$ R-symmetry, indexed by $I=1,2$ with a symmetric metric $\delta^{IJ}$. This doubles the number of supermomenta to $q_{iI}^A$, and superamplitudes now carry factors of $\e^{F_2}$ with
\begin{equation}
\label{susy-factor2}
F_2:=\sum_{i<j} \frac{u^a_{iA}  u_{jBa}}{\sigma_{ij}}\, q^A_{iI}q_j^{BI}-\frac1{2}\sum_{i=1}^n\xi_{iaA}v_{iB}^a\;q_{iI}^Aq_j^{IB}\, .
\end{equation}
Alternatively, we can extend the supersymmetry in a parity invariant way by introducing  supersymmetry generators $Q^\alpha$ of the opposite chirality, leading to supermomenta $q_{iA}$ in the conjugate representation of the tiny group. This leads to  exponential supersymmetry factors $\exp\,{\tilde F_1}$, now built out of  conjugate $\tilde u_{i\dot a}^A$s and $\tilde q_{iA}$s.

The supersymmetry factors $\e^{F_1}$, $\e^{\tilde F_1}$ and $\e^{F_2}$ are supersymmetric under  $Q_\alpha$  by an identical calculation to the 11d case.  Thus, any formula will be supersymmetric if there is no $q$-dependence in the rest of the integrand.

\paragraph{10d formulae.}
We can now introduce 10d formulae that are supersymmetric extensions of the CHY formulae of \cite{Cachazo:2013hca,Cachazo:2014xea}.  In these, gravity amplitudes arise as a double copy of  Yang-Mills amplitudes. Thus our integrands $\CI_n$ in \eqref{CHY} are constructed from reduced Pfaffians or determinants of the CHY matrix $M$ in \eqref{CHY-M} or of the submatrix $A$, as well as the supersymmetric factors $\e^F$,
\begin{center}
\begin{tabular}{ll}
 \text{Super Yang-Mills:}\hspace{10pt} & $\mathrm{PT}(\alpha)\;\pf{}'M\;e^{F_1}$ \\
 Born-Infeld: & $\det{}'A\;\pf{}'M\;e^{F_1}$ \\
 IIASupergravity: & $\det{}'M\;e^{F_1+\tilde{F}_1}$\\
 IIB Supergravity:  & $\det{}'M\;e^{F_2}$\\
 Heterotic Supergravity: \hspace{10pt}&$\det{}'M\;e^{F_1}$.
\end{tabular}
\end{center}
We can also define the Einstein-Yang-Mills superamplitudes of heterotic  supergravity by using the corresponding Einstein-Yang-Mills integrands of \cite{Cachazo:2014xea}. 
All formulae are manifestly supersymmetric, and reduce to the correct bosonic amplitudes.

\paragraph{Factorization.} For CHY-like amplitudes, the scattering equations relate factorization -- a crucial check on any amplitude representation -- to behaviour at the boundary of the moduli space $\widehat{\mathfrak{M}}_{0,n}$ of $n$-points on the Riemann sphere up to Mobius transformations \cite{Dolan:2013isa}:  
\begin{equation}
\partial\widehat{\mathfrak{M}}_{0,n}\simeq \widehat{\mathfrak{M}}_{0,n_\ssL+1}\times \widehat{\mathfrak{M}}_{0,n_\ssR+1}\,.
\end{equation}
The factorization of our formulae here follows very much analogously to the factorization of the analogous 6d formulae as proved in \cite{Albonico:2020mge} and the reader is referred there for full details of  factorization in a closely analogous context. 
Parametrizing the moduli space around this boundary divisor by 
\begin{equation}\label{eq:bdy}
 (x-x_\sL)(\sigma-\sigma_\sR)=\varepsilon\,,\quad \text{with } x\in \Sigma_\sL,\,\sigma\in\Sigma_\sR\,,
\end{equation}
and $\varepsilon\ll 1$,
the polarized scattering equations allow us to introduce spinor data at the junction points $u_{\sR  a\sA}\epsilon_{\sR\alpha}^{\sA}=\sum_{i\in L} u_{ia\sA}\epsilon_{i\alpha}^{\sA}$, such that $\lambda(\sigma)\sqrt{d\sigma}$ descends to the component spheres $\Sigma_{\sL,\sR}$. Moreover, since $\lambda(\sigma)\sqrt{d\sigma}$ is invariant under the inversion \eqref{eq:bdy}, $u_i$ behave as worldsheet spinors of the local bundles at the marked points, $ u_{ia\sA} =i\varepsilon^{1/2}x_{i\sL}^{-1}w_{ia\sA}$. Putting this together, the supersymmetry factors $e^F$ factorize as
\begin{align}
 e^{F_\sN} = \int \!\! d^{4N} \!q_\sL d^{4N} \!q_\sR\, e^{F_\sN^{(\ssL)}+F_\sN^{(\ssR)}}\, G(q_\sL,q_\sR)\,,
\end{align}
where the exponential `gluing factor' $G$ is given by
\begin{equation}
 G(q_\sL,q_\sR) =\det\!{}^{N}\!\big(\epsilon_\sL \epsilon_\sR\big)\,e^{-i (\epsilon_\sL \epsilon_\sR)\!_{\ssA\ssB}^{-1}\,q_\ssL^\ssA q_\ssR^\ssB}\,,
\end{equation}
with $(\epsilon_\sL \epsilon_\sR)^{\sA\sB}=\epsilon_\sL^{a\sA} \epsilon_{\sR a}^\sB $
. This is the correct factorization behaviour for the exponential supersymmetry representation: the exponential in $G$  is dictated by supersymmetry invariance, and the norm ensures agreement with the bosonic sum over states. We have thus verified that all supersymmetric amplitudes  factorize correctly.

\paragraph{Reduction to 4d.}
In the following, we check that our formulae reduce to the correct 4d amplitudes, making contact with the ambitwistor representations \cite{Geyer:2014fka}, which are closely related to the twistor string amplitudes \cite{Witten:2003nn,Berkovits:2004hg,Roiban:2004yf,Witten:2004cp, Geyer:2016nsh}. To implement the reduction, denote the 2-component spinor indices 
by $\Asmall$ and $\dAsmall$, and replace the six-dimensional $\mathrm{SU}(4)$ spinor indices $A,B$ by $I,J=1,\ldots,4$, which will now play the role of $SU(4)$ R-symmetry indices.
In this notation, 10d  spinors decompose in $(4+6)$d to 
\begin{equation}\label{to4d}
\lambda_\alpha=   \left(\lambda_{\Asmall I},\tilde{ \lambda}_{\dAsmall}^I\right)\,.
\end{equation}
The gamma matrices and vectors decompose as
\begin{equation}\label{4dgamma}
\gamma_\mu^{\alpha\beta}\lambda_\alpha\lambda_\beta=(\lambda_{\Asmall I}\tilde{\lambda}_{\dAsmall}^I, \lambda_{\Asmall[I}\lambda^{\Asmall}_{J]}+\half\varepsilon_{IJKL}\tilde{\lambda}_{\dAsmall}^K\tilde{\lambda}^{\dAsmall L})\,.
\end{equation}
For null 4d momenta such as  $k_\mu=(\kappa_{\Asmall}\tilde{\kappa}_{\dAsmall},0)$,
we can perform a global little group normalization $\xi_a=(\xi_I,\xi^I)$ so that  4-momenta $\kappa_{\Asmall}\tilde{\kappa}_{\dAsmall}$ and $\lambda_{\Asmall}\tilde{\lambda}_{\dAsmall}$ give rise to
\begin{equation}\label{eq:lambda-4d}
 \kappa_\alpha ^a = \begin{pmatrix}
 0&\tilde{\kappa}_{\dAsmall}\,\delta_{J}^{ I}\\\kappa_{\Asmall}\,\delta_I^J& 0
 \end{pmatrix},\quad 
  \lambda_\alpha^a(\sigma)=\begin{pmatrix}
 0&\tilde{\lambda}_{\dAsmall}(\sigma)\,\delta_{J}^{I}\\\lambda_{\Asmall}(\sigma)\,\delta_I^J& 0
 \end{pmatrix}.
\end{equation}
 Using $+$ and $-$ to denote  self-dual and anti-self-dual particles respectively, we find that for $+$ the tiny group index can be normalized to be an upper $SU(4)$ index and for $-$ a lower one,
 \begin{subequations}
\begin{align}
  &\epsilon_{ia} ^J = \big(\epsilon\delta_I^J,0\big), \; i\in +\, , && \tilde\epsilon_{ia J}=\big(0,\tilde{\epsilon}\,\delta^I_J\big),\;  i\in -\, ,\\
   &\xi_{iI}^a = \frac1{\epsilon}\big(0,\delta_I^J\big), \; i\in +\, , && \tilde\xi_{i}^{aI}=\frac1{\tilde{\epsilon}}\big(\,\delta^I_J,0\big),\;  i\in -\,.
\end{align}
\end{subequations}
where the prefactors of $\xi$ follow from the normalization condition \eqref{eq:xi_10d} and the scalar $\epsilon$ and $\tilde{\epsilon}$ are all that's left of the polarization data with our choices. 
 These identifications lead to  
 \begin{equation}\label{eq:sols_red_4d}
  u_{iI}^a= 
 u_i(0, \delta^J_I) \, , \;  i\in +\, , \quad
 u_{i}^{aI}=\tilde u_i(\delta_J^I,0) \, , \;  i\in -
 \end{equation}
with identical expressions for $v$ in place of $\xi$ due to the  normalization conditions.  With this,  \eqref{lambda-sigma} reduces to \eqref{eq:lambda-4d} with
$ \lambda_\Asmall$ and $\tilde{\lambda}_\dAsmall$ given by
\begin{equation}
\lambda_\Asmall(\sigma)=\sum_{i\in -}\frac{u_i\epsilon_{i\Asmall}}{\sigma-\sigma_i}\, , \quad 
\tilde\lambda_\dAsmall(\sigma)=\sum_{p\in +}\frac{\tilde u_p\tilde\epsilon_{p\dAsmall}}{\sigma-\sigma_p}\, .
\end{equation}
 and the polarized scattering equations  \eqref{SE10} reduce to 
\begin{equation}
\Big(
\tilde u_i\lambda_{\Asmall}(\sigma_i) , \;u_i\tilde\lambda_\dAsmall(\sigma_i)\Big)= \bigg(
\frac{\kappa_{i\Asmall}}{\tilde\varepsilon_i} , \;\frac{\tilde\kappa_{i\dAsmall}}{\varepsilon_i} \bigg)
\label{10-4SE}\,,
\end{equation}
subject to $(u_p,v_p)=0$ for $p\in +$ and  $(\tilde u_i,\tilde v_i)=0$ for $i\in -$.
These are the familiar 4d refined scattering equations of \cite{Geyer:2014fka} for the N$^{k-2}$MHV sector, where $k$ denotes the number of negative helicity particles.  They have $\left\langle\substack{n-3\\k-2}\right\rangle$ solutions, where $\left\langle\substack{P\\Q}\right\rangle$ denotes the $(P,Q)$ Eulerian number.
 Summing over all sectors,  \eqref{eq:sols_red_4d} incorporates all $(n-3)!$ solutions of the polarised scattering equations.
 
  On the N$^k$MHV sector given by \eqref{eq:sols_red_4d}, the 10d supersymmetry generators reduce  to the familiar 4d generators  $Q_\alpha=(Q_{\Asmall I},\tilde{Q}_{\dAsmall}^{I})$, with
\begin{equation}
(Q_{\Asmall I},\tilde{Q}_{\dAsmall}^{I})=\begin{cases}\big(\epsilon_{i\Asmall}\frac{\partial}{\partial q^I_i},\;\frac{\tilde\kappa_{i\dAsmall}}{\epsilon_i}\,q^I_i \big) & i\in -\\
\big(\frac{\kappa_{i\dAsmall}}{\tilde\epsilon_i}\,q_{iI},\;\tilde\epsilon_{i\dAsmall}\frac{\partial}{\partial q_{iI}} \big) & i\in +
 \end{cases}\,.
\end{equation}
Thus the supermomenta are chiral on the self-dual particles and antichiral on the anti-self-dual particles.
In this MHV sector  we have
\begin{equation}
 \eta_a(\sigma)=\frac{1}{2}\bigg(\sum_{p\in +}\frac{\tilde u_{p}q_{pJ}}{\sigma-\sigma_p},\,\sum_i\frac{u_{i}q_{i}^J}{\sigma-\sigma_i} \bigg)\,,
\end{equation}
and the terms proportional $q_i^Iq_i^J$ in the $F$ vanish due to $\xi_{iaI} v_{iJ}^a=0$. 
The supersymmetry factors then become  $\exp\,F_4^{(\mathrm{4d})}$ for $\mathcal{N}=4$ super Yang-Mills and $\exp\, {F_8^{(\mathrm{4d})}}$ for $\mathcal{N}=8$ supergravity, with $J=1,\dots,\mathcal{N}$ and 
\begin{equation}
 F_{\scalebox{0.6}{$\mathcal{N}$}}^{(\mathrm{4d})}=\sum_{\substack{i\in -\\p\in +}}\frac{u_i\tilde u_p\,q_{i}^Jq_{pJ}}{\sigma_{ip}}\,.
\end{equation}
This is a standard representation for supersymmetry in four dimensions,  known as the link representation \cite{He:2012er}.

 The integrands  can be identified with the 4d integrands of \cite{Geyer:2014fka} after dimensional reduction \cite{Zhang:2016rzb}, with the CHY Pfaffian playing a double role: as the reduced determinant required for the gravity amplitude, as well as the Jacobian from integrating out the $u_i$'s. Thus our formulae reduce correctly to the known 4d formulae.

\section{Discussion}
Much underpinning theory for these equations is likely to follow analogously to that for the polarized scattering equations in 6d \cite{Geyer:2018xgb, Albonico:2020mge} including factorization, BCFW proofs, the existence and uniqueness of solutions,
reductions to other theories in $d<10$, for example to the recent 6d superamplitudes of \cite{Geyer:2018xgb,Heydeman:2017yww, Cachazo:2018hqa, Heydeman:2018dje}. In particular, in higher dimensions we can define an analogue of $H_{ij}$, but each term is now a matrix, $H_{ij}^{ab}=\epsilon_{i\al}^a\epsilon_{j}^{\al b}$ for $i\neq j$, $H_{ii}^{ab}=-e_i\cdot P(\sigma_i)\delta^{ab}$ in 11d.  Thought of as an $n\times n$ matrix with $8\times 8$ matrix entries, it is possible to introduce a reduced quasi-determinants that are equivalent to the CHY reduced determinants and Pfaffians,  we plan to follow up with some of these details in due course  \cite{wip}.  

Other directions include making contact with the semi-pure spinors of  \cite{Berkovits:2002uc}, the pure spinor framework in 10d  and to extend the formulae to include brane degrees of freedom 
\cite{Bandos:2014lja}. The distinction between our constraints and those of \cite{Bandos:2014lja} is that ours are intended to restrict  to pure ambitwistor degrees of freedom, i.e., those of the space of complex null geodesics, whereas, as pointed out   in 
\cite{Bandos:2019zqp}, weaker versions of the constraints might allow one to say something more profound about other M-theory degrees of freedom.

In the 11d and 10d ambitwistor string models, it  remains a key issue to  conduct a full study of the BRST structure of the constraints and associated anomalies. This is of special interest in 10d, where the corresponding RNS model for supergravity is critical. In the spinorial model, we expect that criticality requires a spinorial version of the integrand, similar to the reduced determinant of a matrix $H_{ij}$ forming the integrand in 4d and 6d \cite{Geyer:2018xgb}. We can indeed define an analogue of $H_{ij}$ in $d=10,11$, where each entry is now a matrix, $H_{ij}^{ab}=\epsilon_{i\al}^a\epsilon_{j}^{\al b}$, with the integrand a reduced quasi-determinant, \cite{wip}.


\smallskip

\begin{acknowledgments}
\emph{Acknowledgements:} We would like to thank Nathan Berkovits,   Eduardo Casali, and  Max Guillen  for useful discussions on 11d. YG gratefully acknowledges support from the CUniverse research promotion project ``Toward World-class Fundamental Physics'' of Chulalongkorn University (grant reference CUAASC), as well as support from the National Science Foundation Grant PHY-1606531 and the Association of Members of the Institute for Advanced Study (AMIAS). LJM  is grateful to the EPSRC for support under grant EP/M018911/1.
\end{acknowledgments}

\bibliography{../twistor-bib}
\bibliographystyle{JHEP}

\end{document}